\documentclass[twocolumn,amsmath,amssymb, superscriptaddress]{revtex4}

\usepackage{graphicx}
\usepackage{dcolumn}
\usepackage{bm}
\usepackage{epstopdf}
\usepackage{color}

\begin{document}


\title{Transient trapping of two microparticles interacting with optical tweezers and cavitation bubbles}

\author{Viridiana Carmona-Sosa}%
\affiliation{Instituto de Ciencias Nucleares, Universidad Nacional Aut\'onoma de M\'exico\\ Apartado Postal 70-543, 04510, M\'exico D.F., M\'exico.}
\author{Pedro A. Quinto-Su}%
\email[E-mail: ]{pedro.quinto@nucleares.unam.mx}
\affiliation{Instituto de Ciencias Nucleares, Universidad Nacional Aut\'onoma de M\'exico\\ Apartado Postal 70-543, 04510, M\'exico D.F., M\'exico.}


\begin{abstract}
In this work we show that two absorbing microbeads can briefly share the same optical trap while creating microscopic explosions. Optical forces pull the particles towards the waist of the trapping beam, once a particle reaches the vicinity of the waist, the surrounding liquid is superheated creating an explosion or cavitation bubble that pushes the particle away while lengthening or shortening the trajectories of the surrounding particles. Hence effectively coupling all the trajectories to each cavitation event. 
We find that when two microbeads reach the waist simultaneously within a distance of 2.9$~\mu$m from the beam center in the transverse plane, a larger explosion might result in ejection from the trap. The measured maximum radial displacements $\Delta \rho _c$ due to cavitation are $\Delta \rho _c =3.9\pm 2.2~\mu$m when the particles reach simultaneously with maximum bubble sizes $R_{max}=6.2\pm 3.1~\mu$m, while for individual cases when one of the particles is outside 2.9$~\mu$m prior to cavitation $\Delta \rho _c$ is $2.7\pm 1.2~\mu$m and $R_{max}=4.2\pm 1.6~\mu$m.
We also measure the characteristic timescale of two particle coalescence which is a measure of the expected time that the particles can stay trapped near the waist. The measurements are fitted by a Poisson decaying exponential probability distribution.
A simple one dimensional model shows that the characteristic timescales for transient trapping of multiple absorbing particles decrease as more objects are added. 
\end{abstract}

\maketitle


\section{Introduction}
Highly focused continuous laser beams or optical tweezers have been used to trap microscopic objects that are held by forces that are proportional to the gradient of the intensity \cite{twee1}. 
Observations have shown that multiple particles can be trapped with single optical tweezers \cite{mult1}. 
Furthermore, a focused laser spot can also trap multiple particles at different spatial locations by periodically steering the beam so that the trap is shared \cite{times1, times2, times3}. 
This technique has also been applied for directed diffusion of multiple beads or optical peristalsis \cite{optperistal}.
Other transient trapping regimes periodically block the trapping beam with an optical chopper to study free diffusion while not allowing the particles to drift outside the imaging area. This has been called  blinking optical tweezers \cite{methodsvideo}.

Partially absorbing particles can interact briefly with the beam waist of an optical trap in a cyclic way under the effect of optical forces and microexplosions or cavitation bubbles \cite{steam1}. A bead that is located below the waist of the trapping beam is pulled by gradient forces to the focused spot. However, as the particle approaches the waist the heating rate increases until a small volume of the surrounding liquid is superheated creating a cavitation bubble that expands and collapses in a microsecond timescale. The bubble pushes the particle below the waist where the cycle restarts. Hence the particle is essentially always moving in the direction of the focused spot (millisecond timescale) while the motion due to the explosion lasts a few microseconds. 
%
   \begin{figure*}
   \begin{center}
   \begin{tabular}{c}
   \includegraphics[width=12.5cm]{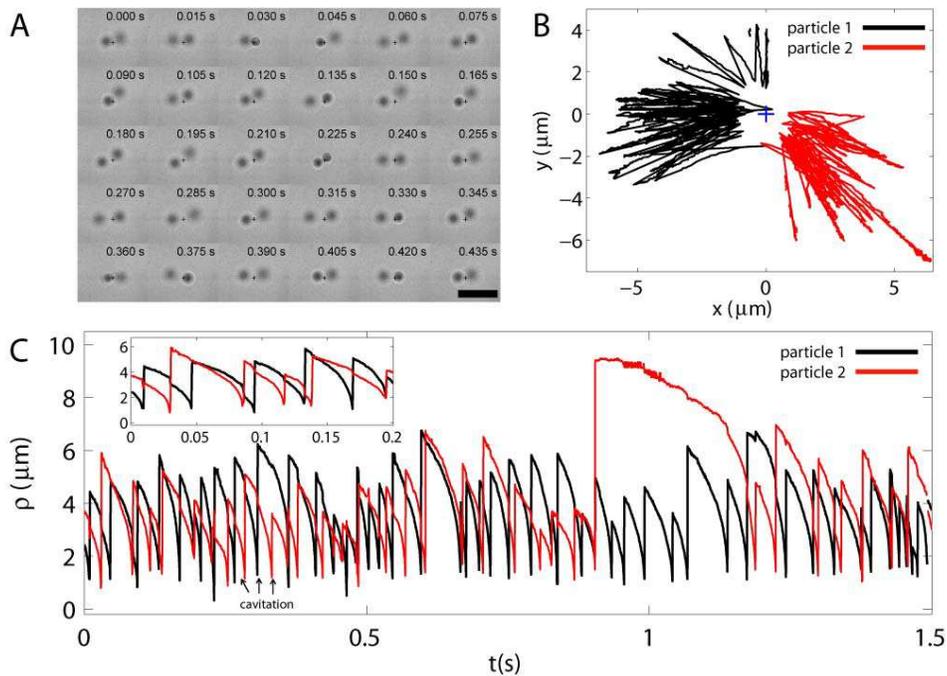}
   \end{tabular}
   \end{center}
   \caption[example] 
   { \label{fig1} 
Two particle interaction. (A) Extracted frames from Video 1. The time step between each frame is $0.015~$s for a total of 0.435 s. The size of each frame is $29\times 24~\mu$m. The scale bar has a width of 15$~\mu$m. (B) Trajectories of the particles in (A) for a time of 1.5 s. The cross marks the position of the trap center. (c) Dynamics for $\rho (t)$ (radial distance from the center of the trap at the $xy$ plane) for both particles. During each cavitation event there is an effect (push, pull, neutral) on the trajectory of the neighboring particle (inset).}
   \end{figure*} 

Here we show that a system of two absorbing microbeads can share a static optical trap. 
Each time a particle reaches the waist creating an explosion, the individual 
cycles of the neighboring particle towards the waist is shortened or lengthened depending on the size of the bubble and the distance to the waist. 
We find that the particles can coexist in the neighborhood of the trap as long as they do not approach the waist simultaneously, which may result in a larger bubble that can eject the particles. 
The characteristic lifetime of the particles in the trap is defined in terms of the time intervals between events of particle coalescence at the waist.  
The distribution of the measured intervals lies in an decaying exponential.

In the case of more than two particles interacting with the optical trap we were able to capture a few events, since the particles are quickly expelled reducing the system to one or two particles. Hence, in those cases the time for particle pair coalescence decreases. 
%
To estimate the decrease in time for a particle pair coalescing in the case of three particles oscillating in a single optical trap we use a simple one dimensional model that only considers the optical force and an empirical model for the interaction between bubble and particles \cite{burb1, transientflow}.    

In the following sections we describe the methods, results, discussion and the one dimensional model.

\section{Methods}
The experiments are done in a near IR optical tweezers setup described in \cite{steam1,burb1}. The trapping laser wavelength is 975 nm and it is focused by a 100$\times$/1.25 NA microscope objective with a transmitted power of 62 mW where single particles can be trapped for a longer time. Larger powers usually result in decrease of trapping time while lower power below 50 mW cavitation is not produced and the particles are not trapped cyclically. The microparticles are magnetic beads (Promag Bangs) with a mean diameter of 3.16$~\mu$m immersed in water (bidistilled water). 
There is one more bead size available from the catalog (Promag Bangs) which has a  $1~\mu$m diameter and can be trapped generating cavitation bubbles for a shorter time (few seconds) than the larger beads \cite{steam1}. Bubble detection for the $1~\mu$m particles requires higher recording speeds (500,000 frames per second) with poorer signal to noise. Hence experiments were only done with the $3~\mu$m beads.
 
The aqueous sample is placed between two microscope coverslides with a separation $\sim 100~\mu$m. 
The dynamics are captured with a high speed video recorder (Photron, SA 1.1) at 2,000 and 300,000 frames per second (fps). The slower recording speed is used to capture the overall dynamics for a few seconds while the higher speed can capture the explosions in a single frame to measure the maximum sizes of the cavitation bubbles \cite{steam1}. 

The position of the particles in the 2,000 fps recordings are only measured in the transverse xy plane, since we could not measure the axial position consistently for all the frames. Furthermore, position detection of two particles in the transverse plane is challenging when the interparticle separation is small and the objects are far away from the geometrical focus of the microscope objective. Here the position is obtained by first subtracting the background and then calculating the absolute value of the digital signal so that the particles (which appear as bright or dark spots depending on height) appear as peaks. The local maximums are detected with the CLEAN algorithm \cite{clean} and used as initial conditions to fit two Gaussians (xy plane). 

The 300,000 fps recording can capture the blurred bubble in a single frame, which is used to measure the maximum bubble size.
The bubble sizes are measured by locating the boundaries of the blurred bubble for a gray level threshold. The threshold is selected \cite{bubblerad} with a value that is 90$\%$ that of the background, this is followed by fitting the located boundary to a circle. The measurement can be used with the Rayleigh formula to estimate the lifetime of the bubbles (few microseconds) which we found has a linear dependence with measurements done with a fast photodiode \cite{burb1} . 

The trapping beam waist is raised above the bottom coverslip at a height of between 15 and 25$~\mu$m. Initially, the particles are near the bottom and within a few micrometers (radial) from the center of the beam waist. 
The particles are pulled to the waist and later are pushed away by the explosion, the total displacement and direction are random as these depend on the size of the bubble and the location where the bubble is created.  Typical displacements are on the order of 10$~\mu$m in the axial $z$ direction and a few microns in the transverse plane $xy$ which is the one that we image \cite{steam1}. 

\section{Results and Discussion}
In Figure 1A we show selected frames from Video 1, recorded at 2,000 fps. 
The particles appear blurred when the axial position is below the waist of the trapping beam. As the particles get closer to the beam waist they appear sharper.  
At 2,000 fps it is not possible to detect the explosions or cavitation bubbles than occur in a timescale of a few microseconds. Hence particle detection is not affected by the presence of the bubbles. The goal is to record the dynamics for several cycles and characterize the interaction between the particles and the typical timescale where the particles can share the same optical trap. 

From the videos we extract the two dimensional position of the particles. The extracted trajectories ($x(t)-x_0$, $y(t)-y_0$) are plotted in Fig. 1B, where ($x_0$, $y_0$) are the coordinates of the trap center (marked with a cross). We observe that as the particles climb towards the trapping beam waist, the distance $\rho (t)= \sqrt{(x(t)-x_0)^2 +(y(t)-y_0)^2}$ in the $xy$ plane to the center of the trap  also decreases. 
Figure 1C shows the dynamics $\rho (t)$ of each particle. The cavitation events are the sharp transitions with relatively large displacements in a single time step. 
In the inset of Fig. 1C we observe that during each cavitation event initiated by the particle nearest to the waist, the neighboring particle is also affected. 
These interactions are similar to those of oscillators that are synchronized through pulses \cite{sync1}. 
In order to locate the cavitation events produced by each particle we use the finite difference vector of the radial position time series $\Delta \rho(t _i) =\rho (t _{i+1}) -\rho (t _i)$ and find the times where the displacement $\Delta \rho$ is larger than 0.72$~\mu$m. We also add the condition that $\rho (t_i) <2.9~\mu$m in order to filter the impulsive displacements of the neighboring particle and just detect the cavitation events produced by the particle interacting with the trapping beam waist.

We also observe that when the particles approach the waist of the trapping beam simultaneously it is likely that one or both are ejected from the trap.  Hence, the criteria of $\rho <2.9~\mu$m that is used to locate cavitation events and filter the effects on neighboring particles is kept to define when the particles coalesce before cavitation (in the xy plane). Hence we define particle coalescence when prior to cavitation both particles are at a distance of $<2.9~\mu$m from the beam center. In order to quantify the displacements $\Delta \rho _c$ after cavitation for the cases of a single particle approaching the beam waist and the simultaneous approach, we measure the displacements for 1637 cavitation events extracted from the 2,000 fps recordings. We found coalescence at 126 events where the displacement $\Delta \rho _c$ is $3.9 \pm 2.2 ~\mu$m, while for the remainder 1511 events $\Delta \rho _c =2.7 \pm 1.2 ~\mu$m. Table I summarizes the results.

\begin{table}[h!]
  \centering
  \caption{Radial displacement measurements. For coalescence events the $\rho $ is the maximum displacement measured for that event. The error is the standard deviation.}
  \label{tab:table1} 
  \begin{tabular}{c c c}
    \hline \hline
  Cavitation & Events   &  $\Delta \rho _c (\mu$m)\\
    \hline
    Individual & 1511 & $2.7 \pm 1.2 $ \\
    Coalescence & 126 & $3.9 \pm 2.2$ \\
\hline
  \end{tabular}
\end{table}

%
   \begin{figure}
   \begin{center}
   \begin{tabular}{c}
   \includegraphics[width=8.0 cm]{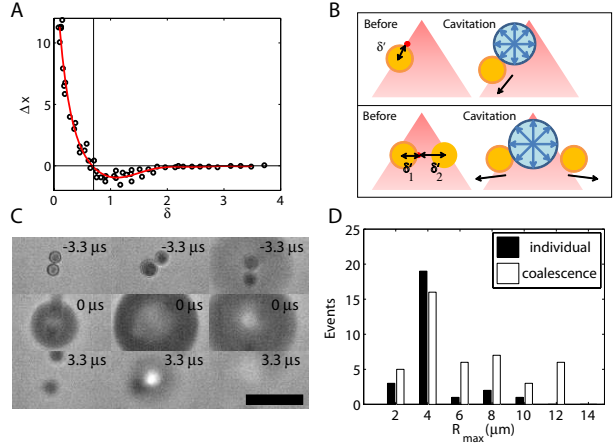}
   \end{tabular}
   \end{center}
   \caption[example] 
   { \label{fig:one} 
(A) Mastercurve for the interaction between particle and bubble. $\delta =\delta '/R_{max}$, $\Delta x =\Delta x'/(2R)$. (B) Bubble particle interaction. Top: single particle. Bottom: two particles reaching the waist simultaneously. 
(C) Frames extracted from recordings at 300,000 fps. Each column is for different events. The width of the scale bar is 15$~\mu$m. (D) Histograms for measured bubbles when there is coalescence according to the definition of $\rho <2.9 ~\mu$m for both particles prior to cavitation and individual events defined as those where one particle is at $\rho >2.9~\mu$m. The bars are centered around a value of $R_{max}$ and represent events a width of $\pm 1~\mu$m around that value.} 
   \end{figure} 
%
%
%

Now we look at the effect of cavitation on the displacement of the particles.
The total displacement on a particle $\Delta x'$ due to a cavitation event depends on the distance $\delta '$ from the center of the particle to the origin of the bubble, the maximum bubble radius $R_{max}$ and the particle radius $R$. The interaction is described by the mastercurve shown in Fig. 2A \cite{transientflow}, where $\Delta x =\Delta x' /(2R)$ and $\delta =\delta '/R_{max}$. A particle can be pushed ($\Delta x >0$) when $\delta <0.7$, pulled ($\Delta x <0$) when $0.7<\delta < 3.7$ or leave the particle in the same position ($\Delta x=0$) when $\delta =0.7 $. In the case of one particle the interaction is in the regime of rejection or pushing $\delta <0.7$.

Figure 2B (top) shows how a bubble is created at the surface of a particle that approaches the waist of the trapping beam, in that case the top surface is the one where higher temperatures are reached (red dot) and as a result a small volume of liquid in contact with that surface is superheated creating a cavitation bubble. Hence the distance from the origin of the bubble to the center of the particle $\delta '$ is the radius of the particle $R$. When two particles are sharing the trap (Fig. 2B bottom), each cavitation event results in a net displacement $\Delta x' _i$ for both particles, the particle that creates the bubble is pushed while the displacement of the neighboring particle depends on the value of $\delta ' _i$ (distance to the origin of the bubble at the surface of the other particle). However the effect on the second particle also depends on the direction in which the first particle is pushed which creates an asymmetry in the flow \cite{steam1}. Hence the mastercurve only provides a qualitative description for the neighboring objects.

The larger displacements observed when the particles approach the waist simultaneously (within 2.9$~\mu$m in the xy plane) could be explained by larger bubbles and by the location where the bubbles are produced. There is the possibility of reaching higher temperatures between the particles (Fig. 2B, bottom) rather than at the top, creating a bubble at the surface of one particle that will push both objects predominantly in the transverse direction. In contrast to the case depicted in Fig. 2B (top) where the displacement of a particle is larger in the axial direction with smaller components in the transverse direction.

In order to measure the maximum bubble sizes $R_{max}$, other experiments were done recording the events at 300,000 fps. Selected frames from three different events are shown in Fig. 2C, where the frames correspond 3.3$~\mu$s before cavitation, during cavitation and $3.3~\mu$s after the explosion. 
The bubble dynamics can be calculated with a Rayleigh-Plesset equation \cite{plesset} while the bubble lifetime $T_{osc} $(time between expansion and collapse) can be estimated using the Rayleigh formula for the collapse time $T_C$ of a spherical bubble\cite{rayleigh} as $T_{osc}=2T_C= 1.82 R_{max} \sqrt{ \frac{\rho _l}{p_0 -p_v} }$, where $\rho _l $ is the density of the liquid (water), $p_0$ the atmospheric pressure (0.1 MPa) and $p_v$ the vapor pressure at room temperature (2330 Pa at 20$^\circ$C ). However deviations are expected for nanobubbles \cite{vogel} and due to non-spherical bubble dynamics \cite{blake}.

The size distribution $R_{max}$ for the bubbles measured is in Fig. 2D. The white bars represent the events where there is coalescence prior to cavitation, that is, when the separation between the particles is less than 5.8$~\mu$m (each within 2.9 $\mu$m from the beam center). The black bars represent events labeled as individual, where for one of the particles $\rho >2.9~\mu$m. In the case of the coalescent events $34\%$ of the bubbles have a size $R_{max}$ larger than 7.5$~\mu$m, which is a significantly larger percentage than for the distribution measured when one of the particles is outside a 2.9$~\mu$m radius around the trapping beam center only $7.7\%$ of the bubbles are larger than 7.5$~\mu$m. For individual events the mean $R_{max}$ is $6.2 \pm 3.1~\mu$m while for the case of coalescence $4.2 \pm 1.6~\mu$m.

These measurements suggest that when both particles are near the waist before cavitation, higher superheat temperatures are reached or larger volumes of liquid are superheated, which should result in higher pressures and a larger bubble originating at the surface of one of the particles \cite{steam1}. Furthermore, the push can have a larger component in the transverse direction as shown in Fig. 2B (bottom).  

In order to characterize the typical times where we would expect to have both particles oscillating in the vicinity of the trap ($\rho < 6~\mu$m) we define a characteristic time as the time interval $\Delta t$ between events where both particles coalesce at the waist (within 2.9$~\mu$m) and are more likely to be ejected or pushed away from the interaction volume of the trap. 
We use the data from the 2,000 fps recordings which include 126 cavitation events where the particles reach the center simultaneously (within 2.9$~\mu$m).
The distribution for the measured $\Delta t$ (between particle coalescence) is shown in Fig. 3A and corresponds to a Poisson process that implies an exponential decay. The continuous line is $\propto \exp -\Delta t /\mu$, where $\mu =0.19~$s. 
The parameter $\mu$ is the expected wait time for a coalescence event where both particles reach a radial distance $\rho <2.9~\mu$m from the beam center prior to cavitation.
The distribution suggests that the characteristic particle coalescence times can be described assuming that the events are independent and with no memory, that $\Delta t$ depends on the random individual cycle frequencies. 
%
%
   \begin{figure}
   \begin{center}
   \begin{tabular}{c}
   \includegraphics[width=7cm]{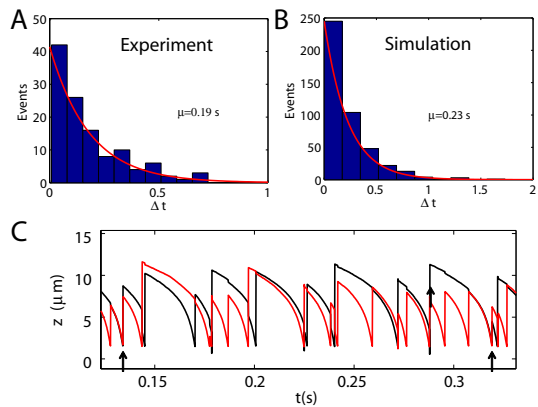}
   \end{tabular}
   \end{center}
   \caption[example] 
   { \label{fig:two} 
(A) Measured distribution for $\Delta t$ (interval between coalescence events) with fitted exponential $\propto \exp (-\Delta t /\mu)$, with $\mu =0.19 ~$s. (B) Distribution for $\Delta t$ extracted from simulated trajectories. $\mu =0.23~$s. 
(C) Simulated trajectories (in $z$) of two particles and their interaction during cavitation.  }
   \end{figure} 

In the next section we use a simple model to simulate a system that yields similar statistics than those measured in the experiment. Then the model is extended for three objects with the same parameters to estimate the expected reduction for the particle pair coalescence time.

\section{Simplified one dimensional model }
A three dimensional model that considers most of the phenomena involved in the experiment is outside the scope of this study. Such a model could include the particle interaction with the trapping light and intermittent changes in the light field after being transmitted by the particles, reflected light \cite{gauthier}, heating of the particle and heat transfer to the liquid, and possible thermal effects like convection currents and thermophoresis that can affect the particle trajectory. Also there is a wide range of maximum temperatures reached before the explosion reported by different experiments \cite{exp1, exp2, exp3, exp4} that depend on many factors like heating rate, liquid purity and heater size. Then there is bubble nucleation \cite{heat1, heat2} and the simulation of the bubble particle interaction.  

In this section we use a simple one dimensional model that includes the bubble-particle interaction (Fig. 2A) that couples the trajectories to simulate particle cycles. All the parameters are adjusted so that the coalescence time (definition also set) near the waist is similar to that from the measured cycles. The goal is to use the model to estimate and quantify the expected reduction of coalescence time for a particle pair in the case of three particles (with the same parameters), since we were not able to measure a sufficient number of events due to rapid particle ejection. 

In the model the particle dynamics are one dimensional in the axial direction along the beam axis so that the particle is always interacting with the trapping Gaussian beam regardless of the distance to the waist. Furthermore, the one dimensional model also removes the random direction in which the particle is pushed during the explosion and the asymmetry created by the particle displacement. We also assume that the presence of an extra particle does not disturb the trapping beam which could be justified by the intermittent cycles when most of the time the particles are far away from the waist, interacting with a small portion of the beam.

In our model the particle dynamics is described with a one dimensional Langevin equation \cite{simul1}
\begin{equation}
m \ddot{z}=F(z)-\gamma \dot{z}+\sqrt{2k_B T\gamma} W(t)
\end{equation}
where $m$ is the effective mass of the particle $\rho _p V_p + \rho _l V_p /2$, with $V_p =4 \pi R^3 /3$ the particle volume, $R$ the radius, $\rho _p$ the particle density and $\rho _l $ the liquid density.  
$\gamma =6 \pi \eta R$ is the Stokes drag with $\eta = 0.001$ Ns$/\mathrm{m}^2$. 
The diffusion coefficient is $D=k_B T/\gamma$ with $k_B$ the Boltzmann constant,  $T=300$K the room temperature and $W(t)$ is the white noise \cite{simul1}.
The force $F(z)$ is proportional to the axial gradient of the intensity of a Gaussian beam $I(\rho, z)=I_0 w_0 ^2 e ^{( \rho ^2/w(z)^2)} /w(z)^2$, where $w_0$ is the radius of the beam waist, $w(z)=w_0(1 +(z/z_R)^2)$, with $z_R$ the Rayleigh range.
Hence the force (with ($\rho =0 $)) is
$F(z)= A_0 z_R ^2 z/(z_R^2 +z^2)^2$, where $A_0$ includes the intensity at the waist $I_0$ and the polarizability of the particle. $A_0$ and the Rayleigh range $z_R$ are chosen so that the trajectories reproduce the measured individual cycle frequencies as a function of the maximum displacements that are correlated with $R_{max}$. 
We choose $z_R= 1.4~\mu$m and $A_0 = 4\times 10^{-15}~$J.
We use the Euler method to solve for $z(t)$ with a time step $dt =10^{-4}~$s. 
In this limit if we divide [Eq. (1)] by $\gamma$, the factor ($m/\gamma$) can be neglected. Hence the non-inertial approximation is \cite{simul1}
\begin{equation}
\dot{z} = F(z)/\gamma +\sqrt{\frac{2k_B T}{\gamma}} W(t)
\end{equation}
The initial condition at $t=0$ s is $z(0)=z_0$. 

In order to simulate the effect of cavitation at the surface of the particle which reaches the beam waist, we impose the condition that when $z$ reaches a value of $z\leq 1.58~\mu$m (at time $t_0$) then a bubble with a maximum size $R_{max}$ is created. 
The bubble size is chosen randomly from a Gaussian size distribution centered at 4$~\mu$m and with a standard deviation of $0.55~\mu$m . 

After choosing $R_{max}$ the new position $z(t_0 +dt)=z(t_0)+\Delta x (\delta)$ of the particle is calculated from the interpolated data \cite{transientflow} in Fig. 2A (red line). 

In the case of two particles that only interact during the cavitation events we added the condition that during cavitation at $t_0$ the distance to the other particle is measured $\Delta z=|z_2(t_0) -z_1(t_0)|$ in order to also calculate the induced displacement (positive, negative or neutral) on that particle. In this way after cavitation (assuming generated by particle 1): $z_1(t_0 +dt)=z_1(t_0)+\Delta x (\delta _1)$, $z_2(t_0 +dt)=z_2(t_0)+ \Delta x (\delta _2) (2R)$, with $\delta _1 = R/R_{max}$ and $\delta _2 = (\Delta z+R)/R_{max}$. 

We define the condition for particle coalescence when both particles are within a distance of $ z_i <3~\mu$m prior to cavitation . The value is chosen to get similar statistics as those for the measured $\Delta t$ in the experiments. 
Figure 3B shows a distribution for $\Delta t$ extracted from simulated trajectories, where $\mu =0.23~$s.  
A sample of the simulated particle dynamics are shown in Fig. 3C. The dynamics are calculated for a million time steps and the events where there is coalescence are marked by the arrows. 
A change in the definition of coalescence will also change the convergence time shown in Fig. 3B.

 \begin{figure}
   \begin{center}
   \begin{tabular}{c}
   \includegraphics[width=9cm]{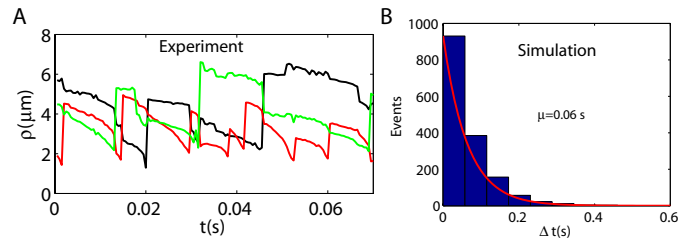}
   \end{tabular}
   \end{center}
   \caption[example] 
   { Three particles oscillating in an optical tweezers. (A) Dynamics $\rho _i(t)$ extracted from Video 2. (B) Two-particle coalescence time $\Delta t$ extracted from simulations with three particles, $\mu =0.063 ~$s.}
   \end{figure} 

\noindent
{\bf Three particles}. 
In our experiments it was difficult to observe three or more particles oscillating within $\rho \sim 6~\mu$m from the center of the trap even when we increased the density of particles in the sample), since in a very short time a pair of particles reach the waist simultaneously ejecting at least one particle from the trap (to a radial distance$\sim 8~\mu$m).
In this way experiments on the interaction between three particles most of the time  are reduced to the interaction between two particles. The presence of an extra particle accelerates the characteristic time for coalescence of a pair $\Delta t$ and typically resulted in the ejection of a single particle at a distance $\rho \sim 8~\mu$m. Hence we only recorded few cycles when three particles were oscillating close to the trap. 

Figure 4A shows the dynamics of three particles extracted from Video 2. 
The simulated characteristic times $\Delta t$ for coalescence of a pair are shown in Fig. 4B. The distribution with $\mu = 0.06~$s (red continuous line) shows that the typical expected times for coalescence are reduced by a factor of four compared with the model for two particles, which explains the shorter time intervals where three particles may oscillate in the vicinity of single optical tweezers. However, The simple one dimensional model neglects thermal effects which could become more important as more particles are added.

\noindent
{\bf{Thermal effects}}.
Now we briefly discuss the thermal effects resulting from the absorption of focused laser beams that have been used in micromanipulation. These effects have been used in static heating configurations where the waist of the focused continuous beam is absorbed by microparticles at the bottom of a substrate of by the substrate.

Studies by Berry and coworkers \cite{berry} where a stable bubble is nucleated due to the absorption of a focused laser beam show that neighboring objects are attracted to the bubble at speeds of several mm/s. They show that Marangoni convection is the dominant effect and explains the strong attractive currents towards the bubble surface and that regular convection has a smaller contribution.

Another thermal effect is thermophoresis \cite{thermo1} where objects are attracted or repelled by the heat source depending on temperature gradients.
In this experiment we could have contributions from normal convection and thermophoresis. However, we did not observe migration of neighboring particles outside the area of the trapping laser. Marangoni currents should not play a role since the bubble lifetime is on the order of a few microseconds. 
We expect that convection and thermophoresis become more important as more than two particles are added, reducing the intermittency of the high temperature source near the beam waist.

\section{Conclusion}
We have shown that two particles can coexist in an optical trap by sharing the beam waist at different times and that the particles may be ejected when reaching the waist  simultaneously (within an area with a radius of 2.9$~\mu$m) due to larger bubbles that result from higher temperatures or a larger superheated volume of liquid. 
The Poisson distribution shows that the system can be modeled as independent random oscillators. 
This is consistent with the observations of the system resetting at each cavitation event erasing the previous trajectories. 
The results are important for applications with microengines, producing micron sized cavitation bubbles with CW laser beams and others that could use impulsive forces at the microscopic scale. This study shows that while it is possible to have multiple particles creating cavitation bubbles around a single focused laser spot, the stability decreases as more particles are added.

\section*{Supplementary Material}
Video 1. Two particles, recording at 2,000 fps, slowed 100 times. 1 second displayed in 100 s at 20 fps. Frame size: $29\times 24 ~\mu$m.\\ 
\noindent
Video 2. Three particles, recording at 2,000 fps, slowed 100 times. 0.5 second displayed in 50 s at 20 fps. Frame size $29\times 24 ~\mu$m.

\section*{Acknowledgments} 
Work partially funded by DGAPA-UNAM project IN104415 and CONACYT National Laboratory project LN260704.


\end{document}